\RequirePackage{amsmath}

\documentclass[runningheads]{llncs}

\usepackage[T1]{fontenc}
\usepackage{graphicx}
\usepackage{hyperref}
\usepackage{color}

\hypersetup{
    colorlinks = true,
    urlcolor = blue,
    linkcolor = blue,
    citecolor = blue
}

\usepackage{amsmath}
\usepackage{amssymb}
\usepackage{amsthm}
\usepackage{booktabs}
\usepackage{color}
\usepackage{enumitem}
\usepackage{latexsym}
\usepackage{makecell}
\usepackage{multirow}

\begin{document}

\title{Tokenizing Stock Prices for Enhanced Multi-Step Forecast and Prediction}

% If the paper title is too long for the running head, you can set an abbreviated paper title here

\author{
Zhuohang Zhu\inst{1}
\orcidID{0009-0009-8362-281X}
\and
Haodong Chen\inst{1}
\orcidID{0000-0003-2254-5629}
\and
Qiang Qu\inst{1}
\orcidID{0000-0002-6648-5050}
\and
Xiaoming Chen\inst{2}
\orcidID{0000-0002-7503-3021}
\and
Vera Chung\inst{1}
\orcidID{0000-0002-3158-9650}
}

\authorrunning{Z. Zhu et al.}

\institute{
University of Sydney, Australia\\
\email{zzhu6520@uni.sydney.edu.au}\\
\email{\{haodong.chen,vincent.qu,vera.chung\}@sydney.edu.au}
\and
Beijing Technology and Business University, China\\
\email{xiaoming.chen@btbu.edu.cn}
}

\maketitle
% typeset the header of the contribution

\begin{abstract}
Effective stock price forecasting (estimating future prices) and prediction (estimating future price changes) are pivotal for investors, regulatory agencies, and policymakers. These tasks enable informed decision-making, risk management, strategic planning, and superior portfolio returns.
Despite their importance, forecasting and prediction are challenging due to the dynamic nature of stock price data, which exhibit significant temporal variations in distribution and statistical properties. Additionally, while both forecasting and prediction targets are derived from the same dataset, their statistical characteristics differ significantly. Forecasting targets typically follow a log-normal distribution, characterized by significant shifts in mean and variance over time, whereas prediction targets adhere to a normal distribution. Furthermore, although multi-step forecasting and prediction offer a broader perspective and richer information compared to single-step approaches, it is much more challenging due to factors such as cumulative errors and long-term temporal variance. As a result, many previous works have tackled either single-step stock price forecasting or prediction instead.
To address these issues, we introduce a novel model, termed Patched Channel Integration
Encoder (PCIE), to tackle both stock price forecasting and prediction.
In this model, we utilize multiple stock channels that cover both historical prices and price changes, and design a novel tokenization method to effectively embed these channels in a cross-channel and temporally efficient manner. Specifically, the tokenization process involves univariate patching and temporal learning with a channel-mixing encoder to reduce cumulative errors. Comprehensive experiments validate that PCIE outperforms current state-of-the-art models in forecast and prediction tasks.


\end{abstract}

\section{Introduction}

% TODO: Too many paragraph breaks, consider, combine multiple paragraphs to one longer paragraph. 
% TODO: Many equations can be combined using \begin/end{gather} 

As of 2023, the worldwide stock market has achieved a valuation of \$109 trillion \cite{2024_stockmk_size}, with the U.S. market contributing an average daily transaction volume of \$110 billion. The ability to forecast and predict stock prices allows investors to make informed decisions about their portfolio positions, such as buying, holding, or selling shares. Furthermore, potential risks and volatilities can be estimated based on stock price forecasts, aiding in the mitigation of losses and minimization of risks. The challenge of stock price forecasting and prediction lies in financial markets' inherently dynamic and often unpredictable nature. The underlying distribution of stock prices is continuously influenced by a myriad of factors, both economic and geopolitical, making accurate prediction particularly difficult. As such, many researches \cite{ACM_19_stock_temporalranking,stock_TRA_OT,stock_num_transformer} have been conducted to tackle this task.

Typically, most of the current works either perform stock price prediction \cite{dva_2024,ACL_18} or stock price forecast \cite{deep_transformer_SPF,KDD17}. Stock price prediction can be defined as predicting the changes of the stock price $\Delta p$ between $t$ and $t+1$, the prediction can be approached as either a numerical estimation or a classification task. In comparison, stock price forecast explicitly aims to forecast the future price $p$ of a stock directly. The effectiveness of each method may vary depending on the specific stock considered. Consequently, both approaches are equally important, as one may outperform the other depending on the stock in question. Although both approaches utilize the same underlying data, the transformation of stock prices from $p$ to $\Delta p$ significantly alters the data's distribution. As a result, models designed for stock price prediction often underperform when applied to stock price forecasting, leading to diminished generalization capabilities.
%1. we proposed a model that works on both way
%2. we proposed a data preprocessing method that improves the performance for many models when performing both tasks.

Currently, a majority of research focuses on a single-step forecast or prediction $\{p_{t+1}\}$ for stock price, instead of making forecast or prediction for the next multiple intervals $\{p_{t+1}, \dots, p_{t+n}\}$. Intuitively, single-step output enables investors to decide whether to buy or sell a stock to secure profits for the following day. On the other hand, predicting stock movements over a longer term offers additional capabilities. Long-term forecast is vital for several applications, including the pricing and hedging of financial derivatives by financial institutions, as well as assessing the risk in the trading books of banks \cite{dva_2024,multi_step_reasons}. On top of that, it gives investors a broader view to make better-informed decisions.

Although multi-step stock price forecast offers enhanced capability when compared with single-step stock price forecast. It also poses significant challenges. The first challenge stems from the intrinsic stochastic nature of stock prices, a phenomenon well-documented in prior research \cite{SPP_ADV_T,dva_2024,socialM_SMP}, which points out stock price typically contains stochastic noises. Additionally, the non-stationary distribution of stock prices, characterized by shifting means and variances over time, exacerbates the difficulty of accurate multi-step forecasts. These problems coupled together make accurate multi-step forecasts more challenging. Moreover, many previous works \cite{iterative_single_step_spp,deep_transformer_SPF} use an iterative method for multi-step forecast, this method is more susceptible to cumulative error as errors in early predictions can propagate and magnify in later predictions \cite{AAAI_dlinear}. To address the aforementioned challenges, we introduce PCIE (Patched Channel Integration Encoder), an encoder model tailored for stock price forecasting.

Additionally, we introduce a data preprocessing technique that simply combines both the stock price $p$ and its change in price $\Delta p$ as inputs for the model. This methodology enhances the performance of our model, as well as that of the baseline models. By incorporating both absolute prices and their fluctuations, our approach provides a more comprehensive representation of market dynamics, which facilitates more accurate forecasts and predictions across various forecasting scenarios.

To demonstrate the effectiveness of PCIE, we conducted comprehensive experiments that demonstrate our model's superior performance in comparison to existing state-of-the-art models with respect to overall forecast and prediction accuracy. Additionally, we assessed the performance enhancements facilitated by our novel data preprocessing method.

The main contributions of this paper are summarized as:

\begin{itemize}
    \item Demonstrate the effectiveness of tokenizing stock price for multi-step stock price forecast and prediction.
    \item We propose a model that achieve \textbf{SOTA} (state-of-the-art) performance for multi-step stock price forecast and prediction
    \item We introduce a novel data preprocessing method that enhances the accuracy of stock price forecasts and predictions.
\end{itemize}

\section{Related Works}

Stock price forecasting and prediction are both popular and challenging tasks, therefore there is a large amount of literature on this topic. In the existing body of literature, we can categorize works into several distinct groups to effectively position our research.

\textbf{Technical and Fundamental Based} Technical-based models aim to forecast future stock prices using quantitative market data, including price and volume. Earlier works focus on using traditional machine learning approaches such as Autoregressive models \cite{lr_stock}, ARIMA models \cite{arima_2014}, Fourier Decomposition \cite{KDD17} and Support Vector Machine(SVM) \cite{svm_2014,arima_svm}. With the rapid advancement of computational capacity and neural network capability, numerous studies have utilized neural networks. This includes using attention-based Long Short-Term Memory (LSTM) \cite{SPP_ADV_T}, Transformer \cite{deep_transformer_SPF} and Graph Neural Network based \cite{GCNN_SPF,graphnn_spp}. In contrast, Fundamental Analysis (FA) employs external data sources, such as news \cite{news_stock}, social media information \cite{socialM_SMP}, earnings call \cite{stock_num_transformer}, or relational knowledge graphs \cite{ACM_19_stock_temporalranking}, to predict price movements. In our research, we concentrate on using Technical Analysis (TA) to assess our methods for processing quantitative financial data.

\textbf{Classification and Regression} For classification models, the objective is to determine whether stock prices will increase or decrease in the subsequent time step, the classification can be either binary \cite{transformer_binary_spp} or multi-class \cite{multi_class_spp}. In comparison, Regression models \cite{dva_2024,deep_transformer_SPF}, the goal is to predict the stock price directly. This provides investors with enhanced information for making informed decisions, such as the ability to purchase the best-performing stocks and use the information in other asset/derivative pricing models. On top of that, most of the regression models either choose to predict the percentage of change in stock price(SPP) \cite{ACL_18}  or forecast the stock price directly(SPF) \cite{KDD17} as these objectives tend to have very different underlying distribution. We aim to address the regression task in our work. On top of that, our model tackles stock price forecast as well as stock price prediction simultaneously.

\textbf{Single and Multiple Steps} Majority of the current works on stock prediction focus on single-step stock price prediction \cite{single_st_transf} and forecast \cite{SPP_DA_KG} for the next time step. On top of that, works such as \cite{iterative_single_step_spp} perform single-step stock price prediction and iteratively apply that to obtain a multi-step prediction. Conversely, research literature that focuses on multi-step forecasting or prediction is rare since accurate multi-step forecasting or prediction is a challenging task. Works such as \cite{dva_2024,KDD17} had attempted to tackle this task. For this work, we will tackle multi-step stock price forecasting as well as prediction since it offers investors more information to make long-term and risk-adjusted decisions.

\section{Methodology}

\subsection{Task Overview}

Multi-step Stock Price Forecast and prediction can be formulated as the following problem: Given an input sequence of X with T trading intervals, $X = \{x_1, x_2, ..., x_T\}$, where $x_t$ at each trading interval t is a vector $C \in \{o_t, h_t, l_t, c_t, \\ v_t, op_t, hp_t, lp_t, cp_t, vp_t\}$. It consists of open price $o_t$, high price $h_t$, low price $l_t$, close price$c_t$, volume traded $v_t$. For $\{op_t, hp_t, lp_t, cp_t, vp_t\} \in C$, each feature is just the percentage change between $t$ and $t-1$ of its original feature, i.e. $op_t = \frac{o_t - o_{t-1}}{o_{t-1}} * 100$.

For stock price forecast task, the target will be the close price with a sequence length of $L$, target sequence = $\{c_1, c_2, ...., c_L\}$. For stock price prediction task, the target will be the close price percentage change with a sequence length of $L$, target sequence = $\{cp_1, cp_2, ...., cp_L\}$.

The data sampling frequency determines the length of the trading interval, it can range from a day when using daily trading data down to 1 millisecond when using high-frequency data.

\subsection{Model Overview}

\begin{figure}[ht]
    \centering
    \caption{Tokenization Process}
    \includegraphics[width=\textwidth]{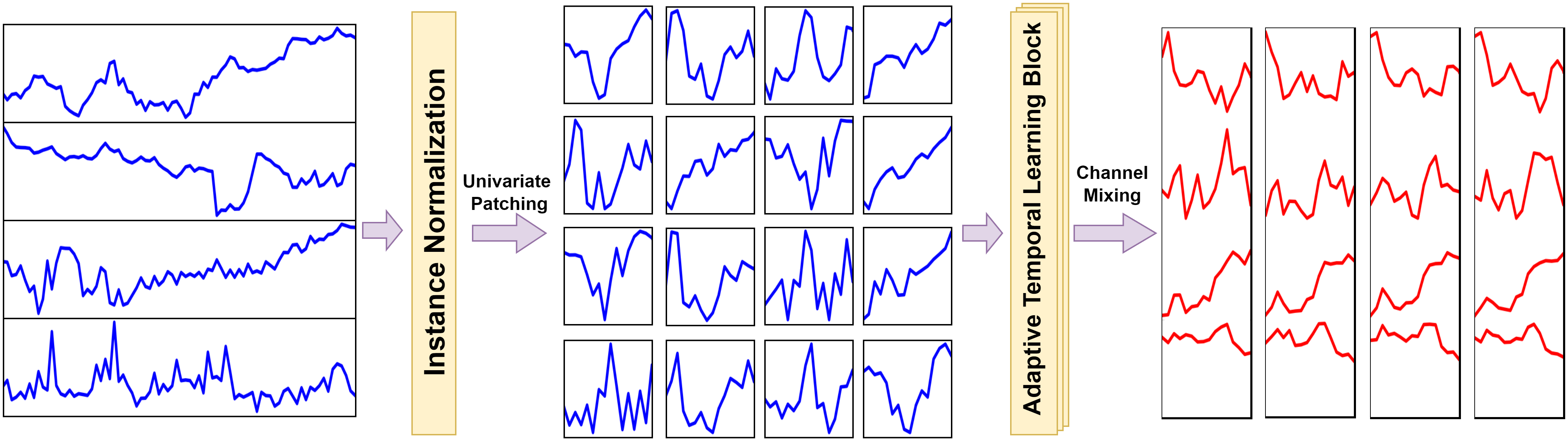}
    \label{fig:patching}
\end{figure}

\begin{figure}[ht]
    \centering
    \caption{PCIE Model Overview}
    \includegraphics[width=\textwidth]{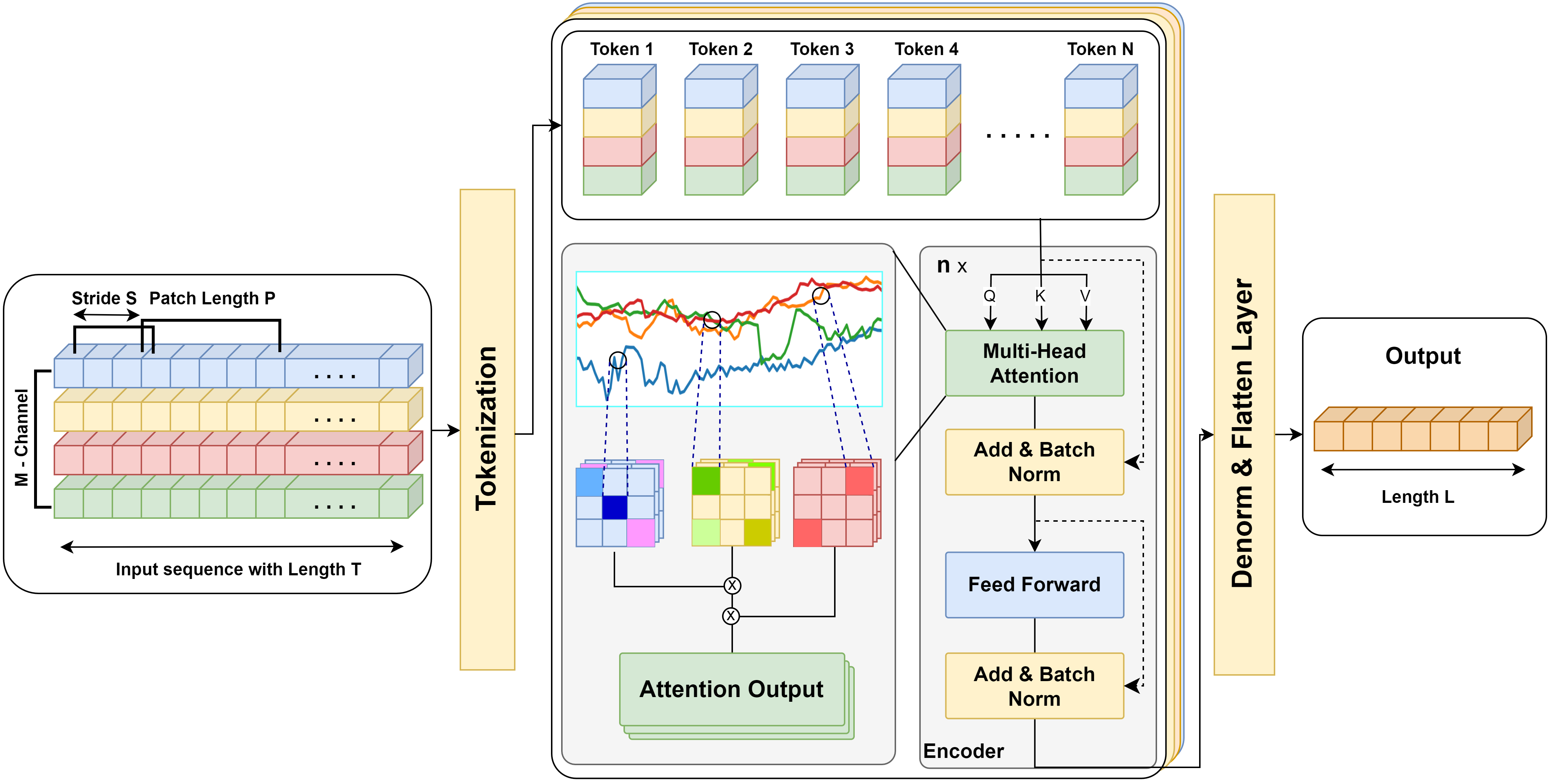}
    \label{fig:model_architecture}
\end{figure}

As shown in Figure \ref{fig:model_architecture}, our method begins by preprocessing the stock price data, dividing it into partially overlapping segments. Each segment is then processed through an adaptive temporal learning module, which maps it into a latent space representation. This process allows the module to focus on the local characteristics of each segment. The transformation (see Figure \ref{fig:patching}) projects local semantic and characteristic information onto the latent representations, enriching the contextual data available for the subsequent self-attention encoder. This structured representation significantly enhances the model's capability to discern relevant patterns within complex financial time series data.

In the subsequent stage, segments from each series are concatenated to facilitate channel mixing. Channel mixing integrates the information across different series, allowing the model to learn correlations across multiple series. This is particularly crucial in stock market analysis, where the target series is often influenced by the movements of other series within the same dataset.

The concatenated vectors are then input into the self-attention mechanism. This model component excels in identifying intricate dependencies between elements in a sequence and extracting patterns in the input sequence. It allows the model to adaptively recognize and respond to significant temporal dynamics and interdependencies between channels.

Lastly, the outputs from the self-attention mechanism are flattened and passed through a linear layer, which obtains the direct multi-step forecasting of stock prices. This method is preferred as it tends to be less prone to cumulative errors in predictions \cite{AAAI_dlinear}, enhancing the reliability of the forecast.

% There are 5 main components in PCIE (Figure \ref{fig:model_architecture}): 1. A patching layer that separates the series and divides each series into patches. 2. A temporal learning block that learns the temporal information between each patch and transforms them into learned vectors. 3. Learned vectors are integrated to form tokens and feed into the self-attention blocks. 4. The Encoder block extracts the temporal information and spatial information of the stock tokens. 5. A linear layer that takes the output from the self-attention blocks and computes the output. Each component will be elaborated with more details below.

%$\{o, h, l, c, v\}$ and treat each series as an univariate series. Then each series is divided into patches.

\subsection{Tokenization Process}

As illustrated in Figure \ref{fig:patching}, the tokenization process includes univariate patching, adaptive temporal learning, and channel mixing. This approach parallels techniques used in natural language processing \cite{kudo2018sentencepiece}, where a word is divided into sub-word tokens and subsequently projected onto embedding vectors, thereby facilitating the model's comprehension of its semantic meaning and how it relates to other words. Similarly, in this context, the tokenization process aids in the effective representation of the dataset. On top of that, it enables the model to better capture the correlation between time steps.

% \begin{itemize}
%     \item Univariate Patching
%     \item Temporal Learning
%     \item Channel Mixing
% \end{itemize}

\subsubsection{Univariate Patching}
\label{subsec:univariate_patching}

Given an input sequence X with a length of L and consisting of M series. Each series is denoted as $X^{(i)}$, as each $X^{(i)}$ is divided into N patches where $x^{(i)}_P \in X^{(i) P \times N}$. Where P is the patch length and S is the stride. There number of patches N is defined by equation \eqref{eq:patch_n}.

\begin{eqnarray}\label{eq:patch_n}
N=\left|\frac{(L-P)}{S}\right|+1
\end{eqnarray}

Depending on the stride S, each patch can be overlapped or non-overlapped. The end of the sequence is padded with the value of the last element of that sequence, with the padding length equal to the stride length.

There are 2 main advantages when using patches. Firstly, it allows the temporal learning block to extract temporal information from the patches within the local scope of $P$. Stock price data typically contains a considerable amount of noise \cite{market_noise}, therefore limiting the scope for temporal learning block will allow it to produce a better-learned representation. Secondly, the number of input vectors is reduced from $L$ to approximately $L/S$. The computational complexity and memory cost can be decreased quadratically by a factor of S. This allows the model to process a longer input sequence which can lead to increase in accuracy and faster training time.

%advantage 1. faster, less computation
%advantage 2. Enable correlation learning block to focus on local sementics. Limited scope sometimes work better. Use an experiment.

\subsubsection{Adaptive Temporal Learning Block}
\label{subsec:adaptive_termporal_learning}

The Adaptive Temporal Learning Block (ATL) is designed to dynamically select the most effective temporal learning methodology from a range of options, including linear, series-independent linear and Multilayer-Perceptrons (MLP) with the objective of producing learned vectors that best represent the underlying patterns for different datasets. The ATL block is updated alongside the model during backpropagation to produce the best representation. It takes an input patch(i.e. vector) with a dimension of $1 \times P$ and outputs a vector with a dimension of $1 \times d\_patch$. This transformation is systematically executed once across all series $\{X^1, ...., X^m\}$ independently. 

% \subsubsection{Linear Model}
The linear model uses a shared linear layer between all the series to encode the representation of each patch. When the underlying distribution and pattern are similar between each series, a better temporal mapping can be learned by using a shared linear model $W_{linear} \in \mathbb{R}^{d\_patch \times P}$.

\begin{eqnarray}\label{eq:linear_model}
X^{(i)}_{d\_patch} = W_{linear} * X^{(i)}_{P} + b_{linear}
\end{eqnarray}

% \subsubsection{Series Independent Linear Model}

The series-independent linear model maps a linear layer to each input series, therefore there are $n$ linear layers for an input sequence with $n$ series. Each linear layer is defined as $W^{(i)}_{linear} \in \mathbb{R}^{d\_patch \times P}$. This method works well when the underlying distribution and patterns between each input series have considerable deviation.

\begin{eqnarray}\label{eq:SIlinear_model}
X^{(i)}_{d\_patch} = W^{(i)}_{linear} * X^{(i)}_{P} + b^{(i)}_{linear}
\end{eqnarray}

% \subsubsection{MLP}

The Multilayer Perceptron (MLP) is employed to transform input patches into learned vectors with higher representational capacity. This choice is crucial for effectively capturing non-linear relationships and complex patterns in stock prices, which are influenced by a multitude of interconnected factors. The added complexity of MLP allows it to produce better representations when the patterns are complex.

\begin{eqnarray}\label{eq:MLP_patching}
X^{(i)}_{d\_patch} = \operatorname{MLP}(X^{(i)}_{P})
\end{eqnarray}

\subsubsection{Channel Mixing}
\label{subsec:channel_mixing}
A channel mixing step will be performed on the output of the ATL block, illustrated by Figure \ref{fig:patching}. This is done by sequentially flattening the output from the ATL block. On top of that, a learnable additive position encoding $W_{pos} \in \mathbb{R}^{d\_model \times N}$ is applied to represent the order of the patches. %determine if use sincos or additive

For an output with a series $\{X^1, ...., X^M\}$, with a position p for the patches. The size of the input vector for self-attention will be $d\_model = d\_patch * M$, illustrated by equation \eqref{eq:flatten_equation}.

\begin{eqnarray}\label{eq:flatten_equation}
X_{d\_model} = \operatorname{Flatten}(X^{(1)}_{d\_patch}, ...., X^{(M)}_{d\_patch}) +  W_{pos}
\end{eqnarray}

\subsection{Channel Mixing Self-attention}

A modified scaled dot-product attention based on \cite{transformer_2017} is used for mapping the learned temporal vectors to the latent representation. For each head $h = \{1, ...., H\}$, 3 attention results matrices $\{Q_h, K_h, V_h\}$ can be obtained by the following equations (\ref{eq:value_m}).

\begin{gather}
Q_h = (X_{d\_model}) W_h^Q  \hspace{15pt}
K_h = (X_{d\_model}) W_h^K  \hspace{15pt}
V_h = (X_{d\_model}) W_h^V \label{eq:value_m}
\end{gather}

The final attention score for each head is calculated by equation \eqref{eq:att_score}, and the value of each head is combined by equation \eqref{eq:multi_head_att}.

\begin{gather}
\operatorname{Attention_h}(Q_h, K_h, V_h)=\operatorname{softmax}\left(\frac{Q_h (K_h)^T}{\sqrt{d_k}}\right) V_h
\label{eq:att_score}
\\
\operatorname{MultiHead}(Q, K, V)=\operatorname{Concat}\left(\operatorname{head}_1, \ldots, \operatorname{head}_{\mathrm{h}}\right) W^O
\label{eq:multi_head_att}
\end{gather}

We substitute the layer normalization in the multi-head attention block with batch normalization as shown in Figure \ref{fig:model_architecture}, since batch normalization shows better performance for time series data \cite{patch_tst,batch_norm_2021T}.

\subsection{Feed-forward Output Layers}

A flatten layer coupled with a linear layer $W_{f} \in \mathbb{R}^{L_f \times d\_model}$ is used to obtain the multi-step stock price forecast with length $L_f$ in a single step, this is also called direct multi-step forecast. 
This can be illustrated by the equation \eqref{eq:ffw_output}. Comparing with other transformer \cite{deep_transformer_SPF,2021_autoformer} based model which uses an iterative forecast method that suffers from cumulative error, direct multi-step forecast is less susceptible to cumulative error \cite{AAAI_dlinear,2021_informer}. 
%cite autoformer, aaai d-linear.
%talk about direct multi step forecast when using a feedforward layer and less cumulative error.

\begin{eqnarray}\label{eq:ffw_output}
X_{L_f} = W_{f} * \operatorname{Flatten}(X_{A}) + b_{linear}
\end{eqnarray}

\subsection{Loss Function}

An MSE loss function is employed to quantify the discrepancy between the predicted value and the ground truth. The overall objective function is \eqref{eq:MSE}.

\begin{eqnarray}\label{eq:MSE}
\text{MSE} = \frac{1}{N} \sum_{i=1}^{N} (y_i - \hat{y}_i)^2
\end{eqnarray}

\subsection{Instance Normalization}

The statistical properties including mean and standard deviation often change over time in Stock price data \cite{SPP_ADV_T,KDD17}. To better address this problem, we utilize an instance normalization technique proposed by \cite{revin_2022}, this technique coupled with our temporal learning block and patching techniques addresses the problem of distribution shift in stock price data.

%forecast
%prediction
%dataset superiority
%use big dataset first, then add

\section{Experiments}

\subsection{Dataset}

We extensively evaluate PCIE and our data preprocessing technique on 2 comprehensive real-world stock datasets, the statistics of the datasets are listed in Table \ref{tab:data_stats}.

\begin{table}[!ht]
\caption{Statistics of the datasets}
\centering
\label{tab:data_stats}
\setlength{\tabcolsep}{0.5em}
{\renewcommand{\arraystretch}{1.2}
\begin{tabular}{|l|c|c|c|}
\hline
Dataset & Duration & $\#$ Stocks & $\#$ Trading Days \\ \hline
US\_71  & 2016/01/04 to 2023/12/29 & 71 & 2011 \\ \hline
US\_14L & 2005/01/04 to 2023/12/29 & 14 & 4780 \\ \hline
\end{tabular}}
\end{table}

For US\_71, it contains the historical price of 71 high trade volume stocks from the U.S. stock market, representing the top 6-9 stocks in capital size and trade volume across the 9 major industries. This approach is similar to other previous works on stock price forecast or prediction \cite{ACL_18,KDD17}. We collect the historical price from 2016/01/04 to 2023/12/29 with a daily sampling interval.

For US\_14L, we selected 14 stocks characterized by high market capitalization and trading volume. We extended the time frame for data collection, capturing historical prices from 2005/01/04 to 2023/12/29, on a daily basis. This longer time horizon was chosen to evaluate the models' performance under conditions of significant distributional shifts, which are common in stock data over extended periods.
Both datasets are collected from Yahoo Finance and anomaly detection is performed to filter out the invalid data. We split all data into training, validation and testing with a ratio of 7:1:2. This is consistent across all models as well as datasets.

%add the details of the second dataset.
%consider adding a table for statistics of the datasets.

\subsection{Baseline Models}
%add arima as the most simple baseline, ez, can do it later

We choose some of the SOTA models as baseline models to demonstrate the capability of our models. On top of that, as the multi-step stock price forecast and prediction tasks have not been widely explored, we also include SOTA models for multivariate long-time series forecasts. The reasoning behind this is that stock price data can be considered a type of time series data.

% \\
% \textbf{ARIMA}
% \\

\textbf{Informer} \cite{2021_informer} is a transformer-style SOTA model for long-time series forecasts that uses a encoder-decoder style structure. It features distilling self-attention which reduces the self-attention output size and computation complexity. It also pads zero on its decoder input to allow direct multi-step forecasts compared to traditional iterative multi-step forecasts.

\textbf{Autoformer} \cite{2021_autoformer} is a transformer-style SOTA model for long-time series forecasts that uses a encoder-decoder style structure. It adds series decomposition and auto-correlation blocks in each attention block. Fast Fourier transform is used in the self-attention to learn the auto-correlation for each series. In addition, a convolution and moving-average based series decomposition is applied to the input data.

\textbf{DVA} \cite{dva_2024} is a diffusion variational autoencoder for stock price prediction. It utilizes a variation autoencoder to generate the sequence prediction and the diffusion process which adds Gaussian noise to the sequence prediction. Finally applying a clean-up block to remove the unwanted noise.

\textbf{PatchTST} \cite{patch_tst} is a SOTA encoder-only transformer for time series forecasts. It performs patching on each series and each series is fed into the encoder individually. However, this approach ignores the dependency between each series.

\subsection{Experiment Settings}

To compare performances for different forecast and prediction lengths, all models are evaluated on a range of different forecast and prediction lengths $L\in\{10,20,40,60\}$. The chosen lengths are typical liquidity horizons required by financial regulators\cite{dva_2024}. We tuned all baseline models according to the recommendations of their corresponding papers for a fair comparison. Our model is tuned according to the MSE loss from the validation dataset. An Adam optimizer is used, the learning rate of is adjusted following the OneCycleLR policy and the maximum learning rate is 0.0001. The batch size is 16. The maximum epoch is 50 with a patience of 19. The patch length $P$ is 4 and the stride $S$ is 1. All the other parameters are tuned according to the characteristics of the dataset. Our experiments are implemented with pytorch 2.1 and cuda 11.8. All of the experiments are done on a NVIDIA GeForce RTX 3090.

\subsection{Results}

The results obtained from our experiments are detailed in Table \ref{tab:result}. The Patched Channel Integration Encoder (PCIE) model has shown superior performance in multi-step stock price forecasting and prediction when compared to the baseline models including PatchTST, Informer, Autoformer, and D-Va. The evaluation metric used across all models is the mean squared error (MSE), complemented by the mean absolute error (MAE) to provide a comprehensive view of the models' accuracy.

The correlation between channels is crucial for stock price data, and models like PatchTST, which convert the input into univariate series, fail to perform well because they omit these essential inter-channel correlations. By treating each channel independently, PatchTST loses valuable information about the relationships between different channels, leading to suboptimal performance in stock price forecasting and prediction.

Models like D-Va and Autoformer perform well on prediction tasks but underperform on forecasting tasks due to their underlying assumptions about the input data distribution. D-Va, for instance, uses a diffusion process that adds noise based on a learned distribution to the target sequence. This approach works well for prediction tasks where the distribution shift is less significant. However, for forecasting tasks, where the distribution of the data can change significantly over time, D-Va's assumptions lead to poorer performance. Similarly, Autoformer employs a learned Fast Fourier Transform (FFT) attention mechanism, which assumes a stationary distribution. This assumption holds less effectively for forecasting tasks, where the data distribution evolves, resulting in subpar performance.

Informer exhibits the worst overall performance among the models evaluated. One contributing factor is the convolution block used to reduce the attention output size, which inadvertently causes information loss. This loss of information is detrimental to the model's ability to accurately capture the complex patterns and dependencies in stock price data, leading to its inferior performance.

By addressing the preservation of inter-channel correlations, adapting to temporal distribution shifts, employing direct multi-step forecasting, and enhancing data representation through tokenization, PCIE overcomes the limitations of existing models and achieves superior performance in both stock price forecasting and prediction tasks.

Furthermore, the performance improvement detailed in Table \ref{tab:data_mix_improvement} demonstrates that our novel data preprocessing method improves the performance across different models.

\begin{table}[t]
\caption{Performance comparison}
\centering
\label{tab:result}
\resizebox{\textwidth}{!}{
\setlength{\tabcolsep}{0.5em}
{\renewcommand{\arraystretch}{1.2}
\begin{tabular}{|cc|cc|cc|cc|cc|cc|}
\hline
\multicolumn{2}{|c|}{Model} & \multicolumn{2}{c|}{\textbf{PCIE}} & \multicolumn{2}{c|}{PatchTST \cite{patch_tst}} & \multicolumn{2}{c|}{D-Va \cite{dva_2024}} & \multicolumn{2}{c|}{Autoformer \cite{2021_autoformer}} & \multicolumn{2}{c|}{Informer \cite{2021_informer}} \\ \hline
\multicolumn{2}{|c|}{Metric} & MSE $\downarrow$ & MAE $\downarrow$ & MSE $\downarrow$ & MAE $\downarrow$ & MSE $\downarrow$ & MAE $\downarrow$ & MSE $\downarrow$ & MAE $\downarrow$ & MSE $\downarrow$ & MAE $\downarrow$ \\ \hline
\multirow{4}{*}{\begin{tabular}[c]{@{}c@{}}US\_71 \\ Forecast\end{tabular}} 
& 10 & \textbf{0.0690} & \textbf{0.1784} & 0.0851 & 0.1903 & 0.2229 & 0.3338 & 0.1292 & 0.2584 & 0.1527 & 0.2904 \\ 
& 20 & \textbf{0.1352} & \textbf{0.2554} & 0.1650 & 0.2985 & 0.2047 & 0.3193 & 0.2112 & 0.3261 & 0.3483 & 0.4271 \\ 
& 40 & \textbf{0.2635} & \textbf{0.3618} & 0.2986 & 0.3987 & 0.3269 & 0.4240 & 0.3134 & 0.4103 & 0.3802 & 0.4613 \\ 
& 60 & \textbf{0.3337} & \textbf{0.4156} & 0.3787 & 0.4496 & 0.4190 & 0.4895 & 0.3897 & 0.4593 & 0.4351 & 0.5010 \\ \hline
\multirow{4}{*}{\begin{tabular}[c]{@{}c@{}}US\_71 \\ Prediction\end{tabular}} 
& 10 & \textbf{1.0027} & \textbf{0.7109} & 1.0660 & 0.7406 & 1.0259 & 0.7278 & 1.0205 & 0.7263 & 1.1650 & 0.7729 \\ 
& 20 & \textbf{0.9961} & \textbf{0.7106} & 1.0570 & 0.7369 & 1.0329 & 0.7296 & 1.0284 & 0.7273 & 1.1484 & 0.7730 \\ 
& 40 & \textbf{0.9793} & \textbf{0.7035} & 1.0272 & 0.7253 & 1.0189 & 0.7124 & 1.0720 & 0.7442 & 1.0878 & 0.7516 \\ 
& 60 & \textbf{0.9983} & \textbf{0.7157} & 1.0138 & 0.7208 & 1.0054 & 0.7243 & 1.0420 & 0.7301 & 1.0882 & 0.7451 \\ \hline
\multirow{4}{*}{\begin{tabular}[c]{@{}c@{}}US\_14L \\ Forecast\end{tabular}} 
& 10 & \textbf{0.1458} & \textbf{0.2590} & 0.1655 & 0.2782 & 0.3472 & 0.4046 & 0.3009 & 0.3881 & 0.2573 & 0.3510 \\ 
& 20 & \textbf{0.2794} & \textbf{0.3625} & 0.2942 & 0.3736 & 0.3893 & 0.4562 & 0.4543 & 0.4789 & 0.3285 & 0.3970 \\ 
& 40 & \textbf{0.5570} & \textbf{0.5203} & 0.5705 & 0.5242 & 0.7245 & 0.6120 & 0.7498 & 0.6275 & 0.7037 & 0.6043 \\ 
& 60 & \textbf{0.8251} & \textbf{0.6355} & 0.8488 & 0.6446 & 0.9461 & 0.7012 & 0.9885 & 0.7248 & 0.9257 & 0.6990 \\ \hline
\multirow{4}{*}{\begin{tabular}[c]{@{}c@{}}US\_14L \\ Prediction\end{tabular}} 
& 10 & \textbf{1.5725} & \textbf{0.8721} & 1.6414 & 0.8845 & 1.5892 & 0.8736 & 1.5920 & 0.8757 & 1.6828 & 0.9140 \\ 
& 20 & \textbf{1.5181} & \textbf{0.8601} & 1.5958 & 0.8804 & 1.5728 & 0.8773 & 1.5457 & 0.8692 & 1.6675 & 0.9004 \\ 
& 40 & \textbf{1.4746} & \textbf{0.8520} & 1.5337 & 0.8681 & 1.5539 & 0.8708 & 1.5078 & 0.8627 & 1.6192 & 0.8801 \\ 
& 60 & \textbf{1.4611} & \textbf{0.8502} & 1.5116 & 0.8639 & 1.5251 & 0.8709 & 1.4906 & 0.8590 & 1.5362 & 0.8714 \\ \hline
\end{tabular}}}
\end{table}

\begin{table}[h]
\caption{Improvement in overall performance when mixing data}
\centering
\label{tab:data_mix_improvement}
\setlength{\tabcolsep}{0.5em}
{\renewcommand{\arraystretch}{1.2}
\begin{tabular}{|l|c|c|c|c|c|}
\hline
  & \textbf{PCIE} & PatchTST & Informer & Autoformer & D-Va \\ \hline
US\_71  & 2.762\% & 2.075\%  & 0.656\%  & 3.307\%  & 1.647\% \\ \hline
US\_14L & 3.985\% & 2.851\%  & 0.907\%  & 2.877\%  & 1.559\% \\ \hline
\end{tabular}}
\end{table}

\subsection{Ablation Study}
%todo
%discuss why tokenization works couple with the characteristics of stock price data
%kw1: ATL able to learn the temporal correlation of each patch, hence helping the self-attention in understanding of the data
%kw2: It also account for the correlation between different channels

The ablation study was conducted to assess the impact of tokenization on the performance of the PCIE model. The experiment involved comparing the PCIE model with and without the tokenization process across two datasets. Results summarized in Table \ref{tab:ablation_study} illustrate the critical role of tokenization in enhancing the model's performance.

\begin{table}[!h]
\caption{Ablation Study}
\centering
\label{tab:ablation_study}
\setlength{\tabcolsep}{0.5em}
{\renewcommand{\arraystretch}{1.2}
\begin{tabular}{|cc|cc|cc|}
\hline
\multicolumn{2}{|c|}{Model} & \multicolumn{2}{c|}{PCIE} & \multicolumn{2}{c|}{\makecell{PCIE \\ (No Tokenization)}} \\ \hline
\multicolumn{2}{|c|}{Metric} & MSE $\downarrow$ & MAE $\downarrow$ & MSE $\downarrow$ & MAE $\downarrow$ \\ \hline
\multirow{4}{*}{\begin{tabular}[c]{@{}c@{}}US\_71 \\ Forecast\end{tabular}} & 10 & \textbf{0.0690} & \textbf{0.1784} & 0.0838 & 0.2004 \\
 & 20 & \textbf{0.1352} & \textbf{0.2554} & 0.1556 & 0.2766 \\
 & 40 & \textbf{0.2635} & \textbf{0.3618} & 0.2964 & 0.3851 \\
 & 60 & \textbf{0.3337} & \textbf{0.4156} & 0.4031 & 0.4458 \\ \hline
\multirow{4}{*}{\begin{tabular}[c]{@{}c@{}}US\_71 \\ Prediction\end{tabular}} & 10 & \textbf{1.0027} & \textbf{0.7109} & 1.1056 & 0.7355 \\
 & 20 & \textbf{0.9961} & \textbf{0.7106} & 1.0793 & 0.7309 \\
 & 40 & \textbf{0.9793} & \textbf{0.7035} & 1.0456 & 0.7208 \\
 & 60 & \textbf{0.9983} & \textbf{0.7157} & 1.1613 & 0.7696 \\ \hline
\multirow{4}{*}{\begin{tabular}[c]{@{}c@{}}US\_14L \\ Forecast\end{tabular}} & 10 & \textbf{0.1458} & \textbf{0.2590} & 0.1712 & 0.2838 \\
 & 20 & \textbf{0.2794} & \textbf{0.3625} & 0.3179 & 0.3896 \\
 & 40 & \textbf{0.5570} & \textbf{0.5203} & 0.6483 & 0.5539 \\
 & 60 & \textbf{0.8251} & \textbf{0.6355} & 0.9455 & 0.6682 \\ \hline
\multirow{4}{*}{\begin{tabular}[c]{@{}c@{}}US\_14L \\ Prediction\end{tabular}} & 10 & \textbf{1.5725} & \textbf{0.8721} & 1.5957 & 0.8828 \\
 & 20 & \textbf{1.5356} & \textbf{0.8683} & 1.5638 & 0.8835 \\
 & 40 & \textbf{1.5129} & \textbf{0.8664} & 1.5536 & 0.8803 \\
 & 60 & \textbf{1.4801} & \textbf{0.8577} & 1.5418 & 0.8786 \\ \hline
\end{tabular}}
\end{table}

The correlation between each channel in stock price data is critical because certain patterns or signals become apparent only when analyzing the relationships between multiple channels within a specific time frame. For example, the correlation between the open price $o_p$ and the low price $l_p$ during the time interval $\in \{t, \dots, t+P\}$  can reveal important market dynamics that are not visible when considering these prices independently.

The process of tokenization effectively captures these inter-channel relationships by producing a representation of the patterns within the time interval $\in \{t, \dots, t+P\}$. Each token represents a segment of data that includes the intricate correlations between different financial indicators over time.

Furthermore, after tokenization, the resultant tokens have a higher dimensionality compared to individual time steps. This increased dimensionality allows for the application of larger attention blocks, which can handle more complex patterns across longer input sequences. The channel mixing attention block leverages this higher-dimensional representation to understand and integrate more sophisticated patterns, facilitating enhanced analysis and prediction across multiple input tokens. This approach ensures that the model comprehensively captures both temporal and spatial dependencies, leading to more accurate and robust stock price forecasting and prediction.

\section{Conclusion and Future Work}

In conclusion, this study introduced the Patched Channel Integration Encoder (PCIE), a novel approach to multi-step stock price forecasting and prediction. The PCIE model incorporates innovative techniques such as univariate patching and adaptive temporal learning, and it has demonstrated state-of-the-art performance in forecasting and predicting stock prices across multiple datasets and prediction intervals.

The ablation study further highlighted the significance of tokenization in enhancing the model's effectiveness, validating our design choices and the synergy between the model components. The integration of absolute prices and their fluctuations through our novel data preprocessing technique has proven to be particularly beneficial, offering a more comprehensive representation of market dynamics.

For future work, several avenues can be explored to extend the capabilities of the PCIE model. Firstly, incorporating additional data types such as macroeconomic indicators or sentiment analysis from news articles could enrich the model inputs and potentially improve predictive performance. Secondly, exploring more sophisticated attention mechanisms could provide deeper insights into the temporal and spatial relationships in financial data. Finally, expanding the model to include online and adaptive learning could make it more robust and responsive to market changes, thereby increasing its practical applicability for dynamic trading strategies.

\begin{credits}
\subsubsection{\ackname} 
Dr. Gregory and Professor Adrian from Optiver provided valuable suggestions and advice for this research.
\end{credits}


\bibliographystyle{splncs04}
\raggedright
\bibliography{references}

\end{document}